\documentclass[review]{elsarticle}

\usepackage{hyperref}
\journal{Ultramicroscopy}

\biboptions{sort&compress}

\begin{document}

\begin{frontmatter}

\title{\textit{ab initio} description of bonding for transmission electron microscopy}

\author[univie]{Jacob Madsen}
\author[emat]{Timothy J. Pennycook}
\author[univie]{Toma Susi\corref{mycorrespondingauthor}}
\cortext[mycorrespondingauthor]{Corresponding author}
\ead{toma.susi@univie.ac.at}

\address[univie]{Faculty of Physics, University of Vienna, Bolzmanngasse 5, 1090 Vienna, Austria}
\address[emat]{EMAT, University of Antwerp, Groenenborgerlaan 171, G.U.431, 2020 Antwerp, Belgium}

\begin{abstract}
The simulation of transmission electron microscopy (TEM) images or diffraction patterns is often required to interpret their contrast and extract specimen features. This is especially true for high-resolution phase-contrast imaging of materials, but electron scattering simulations based on atomistic models are widely used in materials science and structural biology. Since electron scattering is dominated by the nuclear cores, the scattering potential is typically described by the widely applied independent atom model. This approximation is fast and fairly accurate, especially for scanning TEM (STEM) annular dark-field contrast, but it completely neglects valence bonding and its effect on the transmitting electrons. However, an emerging trend in electron microscopy is to use new instrumentation and methods to extract the maximum amount of information from each electron. This is evident in the increasing popularity of techniques such as 4D-STEM combined with ptychography in materials science, and cryogenic microcrystal electron diffraction in structural biology, where subtle differences in the scattering potential may be both measurable and contain additional insights. Thus, there is increasing interest in electron scattering simulations based on electrostatic potentials obtained from first principles, mainly via density functional theory, which was previously mainly required for holography. In this Review, we discuss the motivation and basis for these developments, survey the pioneering work that has been published thus far, and give our outlook for the future. We argue that a physically better justified \textit{ab initio} description of the scattering potential is both useful and viable for an increasing number of systems, and we expect such simulations to steadily gain in popularity and importance.
\end{abstract}


\end{frontmatter}

%\linenumbers

\section{Introduction}
Transmission electron microscopy (TEM) has become an invaluable and versatile tool for materials science and structural biology. Improvements in instrumentation, data analysis methods, and the development of new imaging techniques and detectors continue at a rapid pace, further expanding its capabilities. Simultaneously, retracing the early development of cryogenic electron microscopy, the data obtained from specimens is shifting from being resolution-limited to being dose-limited. This has prompted increasing interest in obtaining the maximum amount of information from each transmitted electron.

In TEM, information about the sample is encoded in changes in the momentum and phase of the electron waves that are scattered by the specimen. At typical kinetic energies of the electron beam, that interaction is dominated by the Coulomb attraction between the negatively charged probe electrons and the positively charged screened nuclei of the atoms of the material, with a weaker contribution arising from the Coulomb repulsion between the probe electrons and the electrons within the sample. The electrostatic potential responsible for the scattering thus includes both the contribution of the screened nuclear cores as well as the valence electron density of the specimen. Due to its very high precision, X-ray crystallography is a powerful technique for the determination of bonding~\cite{contreras-garcia_bonding_2012}. However, due to limitations in x-ray optics, that technique is limited to relatively large sample areas, whereas the higher spatial resolution of electron microscopy makes it suitable for studying bonding at interfaces, edges and defects.

Since nuclear scattering typically dominates in transmission electron microscopy, it is common to ignore the contribution of valence bonding and instead approximate the scattering potential as a superposition of isolated atomic potentials in the so-called independent atom model (IAM). This approximation has been shown to be accurate to within 10\% for low $Z$-number materials~\cite{kirkland_advanced_2010} and is accepted to be within about 5\% in typical cases and for the low scattering angles that are the focus of phase-contrast imaging techniques~\cite{peng_high_2011}.

Furthermore, advanced imaging methods are able to directly reconstruct the electrostatic potential of the sample, whose theoretical description obviously requires valence bonding to be correctly accounted for. Ionic atomic form factors can be used to include bonding effects for compounds with strong ionicity while remaining within the IAM~\cite{Rez1994DiracFockScattering}, providing a much better fit to measurements than similar calculations based on neutral atoms~\cite{Gajdardziska-Josifovska1993AccurateHolograms}. However, for most materials an \emph{ab initio} calculation is the appropriate approach. Even though valence electrons constitute a small contribution to a TEM measurement, they are what fundamentally binds a collection of atoms together into a material and thus are of significant interest.

\subsection{Measuring the electrostatic potential}
There are many different transmission electron microscopy modalities that can be used to access the electrostatic potential. The contrast in high-resolution TEM (HRTEM) is typically produced by the intentional introduction of lens aberrations to alter the relative phases of the different spatial frequencies. The resulting interference causes an image with a contrast depending strongly on the choice of aberrations, but which has been used to detect local changes in charge density~\cite{meyer_experimental_2011} via their influence on scattering.

More commonly, holography is instead used to study local electromagnetic fields. In this technique, a reference wave in vacuum is used to interfere with a wave transmitted through the sample, called the image wave~\cite{dunin-borkowski_electron_2019}. Off-axis holography uses a biprism to interfere the reference and image waves and is presently the most popular electron holography method (though in-line techniques should also be mentioned~\cite{koch_off-axis_2010}). Although the detector recording the resulting interference can only provide an intensity image, the fringes that appear in this intensity provide a way of determining both the phase and amplitude of the image wave. For a thin and light, weakly diffracting sample, the phase is proportional to the projected potential plus a contribution from any magnetic field present, making it possible to measure local variations in both fields from the measured phase. However, the requirement of the reference wave typically limits this to regions of the sample close to vacuum. Holography also requires a high degree of coherence, and background images are needed to remove artifacts such as the influence of imperfections in the positively charged wire used for the biprism.

Multiple modes of phase imaging also exist for scanning TEM (STEM), which offers the added advantage of simultaneously available compositionally sensitive signals such as annular dark-field contrast~\cite{pennycook_scanning_2011}. Although its importance has decreased with the advent of aberration correction, quantitative convergent-beam electron diffraction (CBED) has enabled precise measurements of charge redistribution due to bonding~\cite{zuo_charge_1997,nakashima_bonding_2011}. For resolving electromagnetic fields, differential phase contrast (DPC) has been an increasingly popular choice. The first proposed DPC detector consisted of four quadrants~\cite{dekkers_differential_nodate}, providing modest sensitivity to shifts of the bright field (BF) disk. If the beam is initially illuminating all segments equally, when it encounters an electric or magnetic field with components perpendicular to the beam direction, it will be deflected by the Lorentz force, causing an imbalance in the illumination of the segments. By determining the difference in their signals, one can calculate a deflection vector, and from this, a field vector. This allows DPC to map electric and magnetic fields at medium resolution, for example in devices such as \emph{p-n} junctions~\cite{shibata_imaging_2015}.

With such measurements, one is essentially measuring shifts in the center of ``mass" (CoM) of the BF disk~\cite{shibata_atomic-resolution_2019}. Especially at higher resolutions where the field varies on a scale similar to or smaller than the size of the probe itself, the BF disk does not always shift rigidly~\cite{muller_atomic_2014,close_towards_2015}. In such cases, the redistribution of intensity within the BF disk itself needs to be measured, and it becomes advantageous to have additional segments. This led researchers Shibata and co-workers to develop a 16-segment DPC detector~\cite{shibata_new_2010}, which was used to demonstrate atomic-resolution DPC imaging for the first time~\cite{shibata_differential_2012}, followed later by electric field imaging of atomic columns as well as individual atoms~\cite{shibata_electric_2017,ishikawa_direct_2018}. With the ability to map the local electric field also comes the ability to deduce the charge density via Maxwell's equations~\cite{sanchez-santolino_probing_2018}. The quality of such maps correlates with the ability to determine the CoM of the scattered intensity, and DPC detectors with even more segments are thus beneficial. We note that the University of Tokyo recently developed a 40-segment version of their DPC detector, which starts to resemble a small pixelated detector, albeit with a small number of curved ``pixels".

Fully-fledged pixelated detectors generally provide higher resolution images of scattering and can thus provide an even more accurate determination of the CoM~\cite{muller_atomic_2014,muller-caspary_measurement_2017,argentero_unraveling_2017}. This is particularly true with the development of direct detection cameras that avoid the use of a scintillator to convert electron hits into light~\cite{mir_characterisation_2017,fang_atomic_2019}, making them far more efficient than conventional cameras. Direct electron detection cameras also generally offer far less noisy images, and their speed has been increasing rapidly. Such cameras have spurred interest in so-called 4D-STEM~\cite{ophus_four-dimensional_2019}, in which the 2D scattering intensity is recorded at each probe position in a 2D scan. Speed, which is important because too slow a scan can cause motion blur in the data due to drift, has so far been an advantage of scintillator-based DPC detectors. However, as cameras increase in speed, the higher resolution maps of scattering available from 4D-STEM are increasingly advantageous for a higher-quality determination of the CoM.

4D-STEM provides the ability to computationally recreate any desired imaging mode after taking the data~\cite{pennycook_efficient_2015, hachtel_sub-angstrom_2018}.  
However, it also enables more advanced processing methods such as electron ptychography \cite{hoppe_beugung_1969}, which is a highly efficient means of computational phase imaging that harnesses redundancies available in 4D datasets (see Ref.~\citenum{hue_wave-front_2010} for a discussion of ptychography, as well as Ref.~\citenum{winkler_direct_2020} for a very recent comparison between STEM and off-axis holography). Various ptychographic algorithms exist; iterative methods such as ePIE~\cite{maiden_improved_2009,hue_extended_2011} or mixed-state electron ptychography~\cite{thibault_reconstructing_2013,chen_mixed-state_2020} make use of redundancy contained in diffraction patterns taken from significantly overlapping regions of the sample to constrain the solution. Such iterative methods can be applied both with a defocused or in-focus probe, and often provide the possibility of removing the effect of the probe~\cite{hue_extended_2011} or superresolution~\cite{jiang_electron_2018}.

Direct non-iterative in-focus probe methods that rely on the interference of diffracted convergent beam electron diffraction (CBED) disks to determine the amplitude and relative phases of the transferred spatial frequencies also exist. Such direct methods include the single side-band method~\cite{rodenburg_experimental_1993,pennycook_efficient_2015,yang_efficient_2015,pennycook_high_2019} or Wigner distribution deconvolution method~\cite{rodenburg_theory_1992,yang_simultaneous_2016,yang_electron_2017}, which also offer the possibility of post-acquisition aberration correction~\cite{yang_simultaneous_2016} or superresolution~\cite{nellist_resolution_1995}. Ptychographic imaging also shows greater resilience to temporal incoherence than HRTEM, and does not require vacuum for a reference wave as holography does~\cite{pennycook_high_2019}. In-focus ptychography also provides the benefits of simultaneous $Z$-contrast annular dark-field images, complementing phase images by easing the discrimination of effects due to the total charge density and atomic number. Therefore, with both center of mass and ptychographic imaging benefiting, the development of fast and efficient cameras for 4D-STEM is enhancing our ability to study the local charge density in materials, albeit at the cost of significantly greater data volumes and computational effort.

\subsection{Beyond independent atoms}
These developments underline the increasing need for an accessible and reliable method to describe the full electrostatic potential of real materials for use in transmission electron microscopy scattering simulations. In this Review, we first survey the existing work that has been done towards this end, and then highlight recent exciting developments, especially concerning the use of the projector-augmented wave method of density functional theory instead of earlier more demanding all-electron approaches, making much bigger systems accessible for \emph{ab initio} simulations without compromising accuracy. Finally, we finish with a brief look at what is becoming possible in terms of modeling and the experiment/theory interface in modern transmission electron microscopy.

\section{State of the art}
\subsection{Description of the {\normalfont ab initio} electrostatic potential}
The electron density and thus electrostatic potential of a collection of atoms is described by their quantum mechanical ground-state wave function. However, due to the complicated many-body interactions of the electron states, that wave function cannot be analytically solved except for a few simple molecules. Different approximations have thus been developed to overcome this fundamental difficulty, with notable early work on the lattice potentials and scattering factors of monoatomic crystals conducted within the Hartree-Fock-Slater formalism by Radi~\cite{radi_complex_1970}, and later refined within the Dirac-Fock formalism by Rez and co-workers~\cite{rez_diracfock_1994}.

However, density functional theory (DFT) is unarguably the most popular, widely used and versatile framework for electronic structure theory of molecules and solids~\cite{kohn_nobel_1999}. In DFT, the many-body problem of $N$ electrons with $3N$ spatial coordinates is reduced to a variational solution for the three spatial coordinates of the electron density. This reduction is in principle exact, up to a term that describes electron exchange and correlation that is not analytically known.

While it is possible to solve the ground-state electron density for all electrons, including those in the core levels and in the valence, the numerical description of the orthogonal electron wave functions rapidly oscillating near the nuclei is computationally extremely expensive. Thus, some partition of the treatment of the cores and the valence is typically used to make calculations practical. One of the most popular such approaches, originally developed by Slater~\cite{slater_wave_1937}, is the augmented plane wave (APW) method, where the electron density in non-overlapping volumes around each atom is described by atomic-like functions, which are smoothly connected by plane waves in the interstitial regions.

Most modern implementations have adopted the efficient linearized form of the APW method originally developed by Andersen~\cite{andersen_linear_1975}, where the density near the atoms is described as a linear variation of atomic-like functions. In its so-called full-potential form (FLAPW), no assumptions about the shape of the potential inside the spheres are made (contrary to the early `muffin-tin' approaches). This results in a highly precise description of the all-electron density, including any core relaxation effects. However, while accurate, this method is still computationally highly demanding, limiting its use to a few dozen or at most around one hundred atoms. For specimens of interest to TEM, especially in a realistic slab geometry, this is a severe limitation in capability.

More recently, so-called pseudopotential~\cite{schwerdtfeger_pseudopotential_2011} and projector-augmented wave (PAW) methods~\cite{blochl_projector_1994} have become popular due to their greater computational efficiency. In these approaches, the core electrons are not described explicitly, but rather replaced either by a smooth pseudo-density near the nuclei, or by analytical projector functions that are also smooth in the core region. Both approaches are valid and useful, but the PAW method has the distinct advantage that the real (frozen) core electron density can be analytically recovered by inverting the projector functions. As such, it is arguably the most suitable approach for obtaining efficient and accurate \textit{ab initio} electrostatic potentials for electron scattering simulations.

The use of the PAW approach in electron microscopy simulation was pioneered by Pennington and co-workers who highlighted that the mean inner potentials of group-IV and group-III–V materials were up to 10\% lower for PAW-based DFT calculations than those derived from IAM scattering factors~\cite{pennington_surface_2015}, and in good agreement with FLAPW values found in the literature. This is to be expected: charge rearrangement due to bonding makes the screening charges more delocalized, and thus each core potential has a longer tail, resulting in a lower value of the mean inner potential of the entire crystal. Recently, Susi and co-workers further directly compared the spatial variation of the electrostatic potential obtained using the PAW method to a reference FLAPW calculation, finding them (unsurprisingly) essentially identical~\cite{susi_efficient_2019}.

\begin{table}[t!]
    \centering
    \caption{\textit{ab initio} simulations of the electrostatic potential of materials for TEM. In these works, electron scattering was not modelled, but the measured electrostatic potentials or electric fields were explicitly compared to the simulation. Core electrons have been treated in several ways: FLAPW denotes the accurate but expensive full-potential linearized augmented plane wave all-electron approach; PAW is the projector-augmented wave method, where the exact frozen core orbital density is described by analytic projector functions; and in the pseudopotential approach, the core density is instead replaced by a smooth pseudo-density.}
    \begin{tabular}{c|c|c|c|c}
        DFT code & Core electrons & Material & Refs. \\
        \hline
        \textsc{Castep} & pseudo & Si, Ge, MgO & \citenum{rez_estimates_1995} \\
        \textsc{Wien} & FLAPW & MgO & \citenum{zuo_charge_1997} \\
        \textsc{flapw} & FLAPW & Si, Ge, MgO & \citenum{kim_ab-initio_1998} \\
        \textsc{Wien2k} & FLAPW & Mg & \citenum{friis_magnesium_2003} \\
        \textsc{Wien} & FLAPW & MgB$_2$ & \citenum{wu_valence-electron_2004} \\
        \textsc{Wien2k} & FLAPW & amorphous C & \citenum{schowalter_ab_2005} \\
        \textsc{Wien2k} & FLAPW & Si, Ge, III-V & \citenum{kruse_determination_2006} \\
        \textsc{gpaw} & PAW & films, nanowires & \citenum{pennington_surface_2015} \\
        \textsc{vasp} & PAW & graphene & \citenum{ishikawa_direct_2018} \\
        \textsc{vasp} & PAW & MoS$_2$ & \citenum{Muller-Caspary2018Atomic-scaleMaterials} \\
        \textsc{vasp} & PAW & Al$_2$O$_3$ & \citenum{auslender_measuring_2019} \\
        \textsc{vasp}/\textsc{Elk} & PAW, FLAPW & MoS$_2$, WS$_2$ & \citenum{fang_atomic_2019} \\
        \textsc{vasp} & PAW & SrTiO$_3$, BiFeO$_3$ & \citenum{Gao2019Real-spaceMicroscopy} \\
        \textsc{vasp} & PAW & MoS$_2$, WS$_2$ & \citenum{Wen2019SimultaneousMicroscopy} \\
        \textsc{vasp} & PAW & MoS$_2$ & \citenum{Boureau2020QuantitativeHolography}
    \end{tabular}
    \label{tab:no_scattering}
\end{table}

\begin{table}[b!]
    \centering
    \caption{\textit{ab initio} simulations of the electrostatic potential of materials for explicit TEM electron scattering simulations. See the caption of Table \ref{tab:no_scattering} for a description of the different treatments of core electrons.}
    \begin{tabular}{c|c|c|c}
        DFT code & Core electrons & Material & Refs. \\
        \hline
        \textsc{Wien2k} & FLAPW & MgO & \citenum{deng_charge_2007} \\
        \textsc{Wien2k} & FLAPW & graphene, hBN & \citenum{meyer_experimental_2011,kurasch_simulation_2011} \\
        \textsc{Siesta} & pseudo & graphene & \citenum{wang_efficient_2013} \\
        \textsc{Quantum Espresso} & pseudo & AlN & \citenum{Odlyzko2016AtomicPredictionsb} \\
        \textsc{exciting} & FLAPW & graphene & \citenum{Pardini2016MappingTheory} \\
        \textsc{abinit}/\textsc{Elk} & pseudo & WSe$_2$ & \citenum{borghardt_quantitative_2017, winkler_absolute_2018} \\
        \textsc{Wien2k} & FLAPW & GaN & \citenum{muller-caspary_measurement_2017} \\
        \textsc{vasp} & pseudo & SrTiO$_3$ & \citenum{Oxley2018AccurateTheory.} \\
        \textsc{gpaw} & PAW & graphene, hBN & \citenum{susi_efficient_2019,madsen_abtem_2020} \\
        \textsc{castep}/\textsc{Wien2k} & pseudo, FLAPW & hBN & \citenum{martinez2019direct}\\
        \textsc{fplo-18} & FLAPW & graphene, hBN & \citenum{kern_autocorrected_2020} \\
        \textsc{gpaw} & PAW & GaP & \citenum{heimes_measuring_2020}
    \end{tabular}
    \label{tab:scattering}
\end{table}

In Tables~\ref{tab:no_scattering} and \ref{tab:scattering}, we list published studies that have used DFT to calculate the electrostatic potential of a specimen in the context of transmission electron microscopy. We have chosen to limit ourselves to studies that have explicitly compared the potential or field itself, or a resulting image or diffraction pattern, to an experimental measurement, leaving out the numerous works where DFT has been used to model any specimen properties (such as strain) that indirectly influence image contrast. In many cases, only the mean inner potential (MIP) --- the zeroth-order term in the Fourier expansion of the full Coulomb potential --- which depends on the local composition, density and ionicity of the material was computed, as this was sufficient to describe the experimental results, and no explicit electron scattering simulation was made (Table~\ref{tab:no_scattering}). In other studies, image simulations were performed based on the DFT potentials to go beyond the independent atom model (Table~\ref{tab:scattering}).

Since scattering simulations and 4D-STEM are explicitly real-space methods, there are some advantages, at least in terms of simplicity, in using a real-space DFT code. Furthermore, the description of vacuum regions that are required for TEM simulations in the typical slab geometry is as computationally expensive as that of the interstitial regions between atoms when using plane waves, and Fourier transforms are required to recover the real-space electron densities. Of the methods that have been used to date, most FLAPW and \textsc{vasp}-based PAW approaches use reciprocal-space plane-wave bases for the wave functions. Instead, as arguably the most popular PAW-based real-space code~\cite{mortensen_real-space_2005,enkovaara_electronic_2010}, \textsc{gpaw} is ideally suited for describing the scattering potential for TEM simulations. Furthermore, its excellent parallel scaling~\cite{enkovaara_electronic_2010} and the option of using efficient localized basis sets~\cite{larsen_localized_2009} allow large systems with vacuum to be effectively treated.

\subsection{Beyond the kinematical approximation with static atoms}

The kinematical (diffraction) approximation assumes that each electron scatters only one time in the specimen and only elastically. This holds when a specimen is thin and weakly interacting enough, allowing for analytical treatments of scattering. However, in general specimens, not only is this approximation violated and multiple dynamical scattering occurs, but the electrons also undergo inelastic energy loss which results in a loss of beam coherence and contributes to a 'background' signal of inelastically scattered electrons. Furthermore, the atoms of the target are not, in reality, static but vibrating due to zero-point and thermal phonon occupations. These effects can all be described by modern electron propagation simulations and can in principle be incorporated into an \emph{ab initio} framework. There is not yet a universally agreed upon, computationally tractable and theoretically robust approach to including thermal vibrations in a DFT-based TEM image simulation.

\subsubsection{Inelastic scattering}

While spectroscopy as such is beyond our scope here, it should be acknowledged that STEM-based electron energy-loss spectroscopy (EELS), which relies on the analysis of inelastically scattered electrons, has emerged as an immensely powerful characterization tool down to the level of single atoms~\cite{suenaga_atom-by-atom_2010,zhou_atomically_2012,ramasse_probing_2013,nicholls_probing_2013,susi_single-atom_2017}. This technique has become ever more capable with the advent of modern electron monochromators~\cite{krivanek_vibrational_2014}, and the modeling of EELS signals is a vibrant and important research field~\cite{kapetanakis_low-loss_2015,senga_position_2019, hage_single-atom_2020, zeiger_efficient_2020}. However, inelastic scattering is also important to include in quantitative image simulation of thicker samples~\cite{Forbes2011ThermalMicroscopy}, although studies using \emph{ab initio} potentials have so far mostly neglected it.

Inelastic scattering can be modeled via an absorptive potential, where the imaginary part of a complex electrostatic potential is used to describe the loss of electrons from the elastic channel. Absorptive form factors have been developed for the case inelastic scattering due to phonon excitations~\cite{Bird1990AbsorptiveDiffraction}. Although this is computationally efficient, the method's serious weaknesses are that the electron flux is not conserved, and high-angle scattering is underestimated~\cite{Forbes2011ThermalMicroscopy}. For the case of phonon excitations, this is largely solved by the frozen phonon approximation as described in a section below.

For typical thin-foil specimens, due to their relatively high energies and cross sections, bulk plasmons are the most prominent energy-loss channel~\cite{egerton_electron_2009}. Thus, a fully accurate description of electron propagation would also need to account for plasmons, which can be effectively treated by \emph{ab initio} approaches including linear response theory and real-time time-dependent DFT~\cite{varas_quantum_2016}. While it has become standard to include phonon scattering in simulations, the inclusion of plasmon scattering have been largely neglected, perhaps because they are less important for high-angle scattering~\cite{Beyer2020InfluenceMicroscopy}. However, with the increased focus on techniques that include low-angle scattering this may be changing, and two distinct methods of including plasmon scattering have recently been reported. Both work together with the multislice algorithm (see Section~\ref{sec:simulation} below); however, one method uses a transition potential~\cite{Beyer2020InfluenceMicroscopy} while the other relies on the Monte Carlo method~\cite{Mendis2019AnLosses}.

Inner-shell ionization is of particular interest as a spectroscopic signal. EELS simulations, including dynamical scattering, rely on combining multislice simulations with inelastic scattering cross sections for the elements of interest. Several codes implement this method (see Table~\ref{tab:simulation}), and there have been recent efforts to improve its computational scalability~\cite{Brown2019AScattering}. These approaches typically use atomic inner-shell ionization cross sections, which fails to incorporate solid-state effects that give rise to the energy-loss near-edge structure (ELNES). A limited number of studies go further by combining density functional theory and dynamical electron scattering for the calculation of spatially resolved STEM ELNES~\cite{witte_eels_2009, prange_eels_2012, oxley_eels_2014}. These show that certain features of observed spectra can only be simulated by taking both the local environment and experimental conditions into account.

Since both inelastic energy loss and multiple (dynamical) scattering are most pronounced for thicker samples, these effects are often treated together in more advanced multislice simulations, as we will discuss below.

\subsubsection{Dynamical scattering}

Studies using \textit{ab initio} image simulation have so far mostly looked at 2D materials, where neither inelastic or multiple scattering plays an important role. However, that is not due to any fundamental limitation, as for example \textsc{gpaw} can handle thousands of atoms, even when including an explicit description of van der Waals interactions~\cite{larsen_libvdwxc:_2017}. However, 2D materials do require less computational effort, and since the projection problem is eliminated and dynamical scattering is negligible, measurements in 2D materials are easier to interpret.

A challenge in recovering electric fields for samples thicker than a few layers is that dynamical scattering deteriorates the reconstructed electrostatic potential~\cite{close_towards_2015, Muller-Caspary2017MeasurementMicroscopy}. For example, the first moment of diffraction patterns becomes difficult to relate to the electric fields when dynamical scattering is present, as it does not even vary monotonically with the increasing projected potential~\cite{close_towards_2015, winkler_direct_2020}. The multislice algorithm by its nature includes dynamical scattering, hence comparison to simulation has been and remains necessary for interpreting and validating experiments. 

Charge transfer measurements due to bonding in bulk crystals using TEM have been demonstrated, including the aforementioned studies on measuring the mean inner potential. Measurements of the structure factors of crystals using quantitative CBED have been shown to be accurate enough to measure charge transfer in more detail~\cite{nakashima_bonding_2011, Pennington2018Neural-network-basedData}. In particular, Nakashima showed that fitting a CBED pattern to a full dynamical theory allowed them to determine the structure factors of aluminium with enough accuracy to resolve the bonds~\cite{nakashima_bonding_2011}.

The general problem of reconstructing the 3D potential from TEM measurements is a difficult one. Some recent works have begun to address this problem~\cite{VanDenBroek2012MethodScattering, yang_simultaneous_2016, Gao2017ElectronImaging, Pelz2020ReconstructingAlone, chen2021electron}, but it will likely require more effort before such methods reach the required fidelity for measuring bonding.

\subsubsection{Effect of thermal vibrations}
Atoms in real materials are not static, even at zero temperature. Thermal diffuse scattering (TDS), or electron-phonon scattering, is thus important both in STEM and HTREM~\cite{Wang2003ThermalApproaches, Forbes2011ThermalMicroscopy}, and is responsible for features including diffuse backgrounds and Kikuchi lines as well as for a large part of the annular dark field  signal originating from phonon scattering~\cite{hage_phonon_2019}.

The effect of TDS is usually unavoidable in STEM performed at room temperature, even when using contemporary energy filters. However, some TEM imaging modes are much less sensitive to TDS than others. It has been shown that the first moment of diffraction patterns is relatively unaffected for a wide range of specimen thicknesses~\cite{winkler_direct_2020}, hence expensive calculations of TDS may be unnecessary for comparison to such measurements. Electron holography, on the other hand, is a perfect filter for TDS, since inelastically scattered electrons do not contribute to the coherent sideband~\cite{winkler_direct_2020}.

While TDS is commonly used in conjunction with the IAM, image simulations relying on \emph{ab initio} potentials have until now only included TDS using limited or post hoc approaches. Quantitative comparisons are scarce, but Susi found that while \emph{ab initio} simulations of graphene and hBN electron diffraction intensities were qualitatively correct (unlike the IAM), they slightly overestimated the relative intensities of higher diffraction orders~\cite{susi_efficient_2019}. The calculation was brought into an excellent agreement with the experiment by approximating TDS using a post-hoc multiplicative Debye-Waller factor. Strictly speaking this is only applicable to monoatomic materials with a single atomic coordinate in the symmetry-reduced unit cell, such as graphene, but appears to also be a good approximation for hBN.

The most common and successful method for simulating TDS with dynamical scattering is the so-called frozen phonon model~\cite{VanDyck2009IsScattering}. The basic idea relies on a rather classical model in which each electron sees a different configuration of atoms displaced from equilibrium by thermal vibrations. The quantum excitation of phonons (QEP) model provides a physically better founded description of the processes behind TDS~\cite{Forbes2010}. Nonetheless, in the statistical limit, the frozen phonon approximation has been shown to give the same results as the QEP model and its simplicity has lead to its widespread use. Molecular dynamics simulations can generate a thermal ensemble of atomic structures~\cite{Krause2018UsingAlGaN}; however, the frozen phonon structures are instead usually created by independently displacing each atom according to a Gaussian distribution, i.e., the Einstein model. While the Einstein model describes high-angle scattering well, it has been shown that subtle features at low scattering angles are not reproduced~\cite{Muller2001SimulationCurve}.

There are no fundamental issues for combining the frozen phonon model with \textit{ab initio} image simulation. The thermal ensemble can be derived from the Einstein model, classical or \textit{ab initio} molecular dynamics, or from calculated phonon dispersion curves~\cite{Muller2001SimulationCurve}. However, simulating the potential of up to a few hundred independent atomic structures, required by both the frozen phonon and Born–Oppenheimer models, will in many cases be computationally too costly. Inclusion of TDS may also require a much larger unit cell than would otherwise be needed to represent a reasonably sized thermal ensemble. Oxley suggested a possible workaround in which only the contribution to the potential from nuclear and core electron charges are shifted and thus only requiring one DFT simulation~\cite{Oxley2018AccurateTheory.b}. However, we have not found any rigorous evaluation of the accuracy of this approach in the literature yet.

There are methods of including TDS, expanding on the absorptive potential approach, that do not involve a large number of atomic configurations~\cite{Croitoru2006AnImages, Rosenauer2008AnMicroscopy}. These have generally been replaced by the frozen phonon approach, probably due to improved computer hardware, but it may be worth re-evaluating these methods in the context of \emph{ab initio} image simulations due to the benefit of improved efficiency. In general, it is our view that this area would greatly benefit from further work to evaluate the accuracy and cost of different methods of simulating TDS with DFT potentials.

\subsection{Simulation of electron scattering}\label{sec:simulation}
Simulations are often required to interpret measurements obtained via different TEM imaging modalities, or to test ideas for a study. Notably, a significant number of 4D-STEM works rely heavily on simulations~\cite{Liu2013, LeBeau2010, Yang2017, Oxley2020}. These are typically performed using either Bloch wave calculations~\cite{Humphreys1979, Bethe1928}, or the multislice algorithm~\cite{Cowley1957, Goodman1974}. Bloch wave methods do not scale favorably with system size, and thus 4D-STEM simulation studies usually employ the multislice method. This consists of two main steps: first, the electrostatic potential of all atoms is calculated and distributed into a series of 2D slices; and second, the electron wave is initialized and propagated through the potential slices. Depending on how the propagated wave is analyzed in real or reciprocal space, either images or diffraction patterns can be easily simulated.

In several published works that have used the \textit{ab initio} approach, only a single propagation through a projected 2D potential within the projection approximation (PA) was performed~\cite{meyer_experimental_2011,kurasch_simulation_2011,wang_efficient_2013,Pardini2016MappingTheory,kern_autocorrected_2020}. Although this is a good approximation for very thin and light specimens, even for truly 2D materials composed of light elements, there seem to be small quantitative differences between the PA and a multislice simulation~\cite{susi_efficient_2019}. For thicker specimens, the projected potential is obviously insufficient for the description of dynamical diffraction. Accordingly, several recent works have performed full multislice simulations based on a 3D \textit{ab initio} electrostatic potential~\cite{deng_charge_2007, Odlyzko2016AtomicExperiments,Muller-Caspary2017MeasurementMicroscopy,borghardt_quantitative_2017,Oxley2018AccurateTheory.,susi_efficient_2019,madsen_abtem_2020,martinez2019direct,heimes_measuring_2020}.

The multislice method is quite efficient for plane-wave or single-probe diffraction simulations, but STEM experiments may record images with thousands or even millions of probe positions. Large 4D-STEM simulations thus require parallelization over multiple central processing units (CPUs) or the use of one or more graphics processing units (GPUs). The most recent developement is the PRISM algorithm~\cite{Ophus2017}, which offers substantial speed-up for such simulations.

There are a large number of codes implementing multislice simulations with varying degrees of support and features; however, not all of these can output the full 4D signal (probe position dependent CBED). In Table~\ref{tab:simulation}, we list some prominent open-source codes that can model 4D-STEM signals. All of the listed codes currently use the IAM to calculate the electrostatic potential, apart from the recently released \textsc{abTEM} code~\cite{madsen_abtem_2020}, which was built to directly support simulations with \textit{ab initio} potentials through integration with \textsc{gpaw}. Heimes and co-workers also recently reported on the latter's upcoming integration with the \textsc{STEMsalabim} code~\cite{heimes_measuring_2020}.

Simulation codes have typically been written in C++ or Fortran for high numerical performance but are somewhat rigid concerning the types of readily available simulation modes. The rapid recent growth of novel imaging techniques calls for flexible simulation tools that are easy to develop and adapt. Some notable codes offer a degree of flexibility through a scripting interface using Python or Matlab~\citenum{Pryor2017, Lobato2015}. The Python language has been gaining significant popularity in scientific computing in recent years, as it provides a low barrier to entry, easy code modification due to being an interpreted programming language, and an extensive list of open source projects to draw on.

\begin{table}[t!]
    \centering
    \caption{Simulation codes that output the full four-dimensional scanning transmission electron microscopy (4D-STEM) signal that are open source and currently maintained.}
    \begin{tabular}{c|c|c|c}
        Simulation code & Language & Features & Refs. \\
        \hline
        \textsc{EMSoft} & Fortran & GPU & \citenum{Graef2019} \\
        \textsc{$\mu$STEM} & Fortran & GPU, EELS & \citenum{Allen2015} \\
        \textsc{MULTEM} & C++, Matlab & GPU, EELS & \citenum{Lobato2015} \\
        \textsc{STEMsalabim} & C++ & Multi CPU & \citenum{Oelerich2017} \\
        \textsc{Prismatic} & C++ & Multi CPU/GPU, PRISM & \citenum{Pryor2017} \\
        \textsc{Dr. Probe} & C++ & GPU & \citenum{Barthel2018} \\
        \textsc{abTEM} & Python & GPU, PRISM, \textit{ab initio} & \citenum{madsen_abtem_2020} \\
        \textsc{py\_multislice} & Python & GPU, PRISM, EELS & \citenum{brown_python_2020}
    \end{tabular}
    \label{tab:simulation}
\end{table}

Two recent codes, \textsc{abTEM}~\cite{madsen_abtem_2020} and \textsc{py\_multislice}~\cite{brown_python_2020} are implemented in pure Python, which arguably offers an easier path for extending and integrating them with other Python tools. In these codes high performance is retained via the use of modern CPU- and GPU-accelerated software libraries such as PyTorch~\cite{NEURIPS2019_9015}, CuPy~\cite{cupy_learningsys2017}, and Numba~\cite{numba}. Other important open source projects for TEM include the Atomic Simulation Environment~\cite{larsen_atomic_2017} for building atomic structures and interacting with atomistic simulation codes that are integral to \textsc{abTEM}, and \textsc{py4DSTEM}~\cite{Savitzky2019}, which is being developed for the analysis of 4D-STEM data and includes integration into the open-source Python electron microscopy software, Nion Swift.

\subsection{Experimental works}
\begin{figure}
    \centering
    \includegraphics[width=0.7\textwidth]{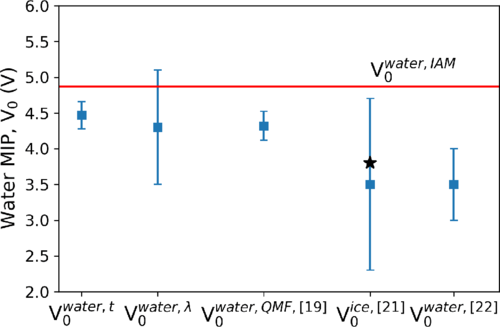}
    \caption{Comparison of mean inner potential (MIP) values of water from several experimental measurements. Red solid line corresponds to an independent atom model (IAM) calculation. For detail and the cited references, please see the original article. Reproduced with permission from Ref.~\citenum{yesibolati_mean_2020}, Copyright \copyright\ (2020), American Physical Society.}
    \label{fig:water}
\end{figure}

Finally, we wish to highlight some illustrative recent examples that clearly demonstrate how going beyond the IAM has been vital for describing experimental data in a quantitatively correct manner. Most commonly thus far, the mean inner potential of materials has been measured with electron holography and compared to an \emph{ab initio} calculation (see Table~\ref{tab:no_scattering}). After pioneering work by Kim on Si, Ge and MgO~\cite{kim_ab-initio_1998}, several groups have extended such studies to other semiconductors, borides, oxides and carbon materials~\cite{friis_magnesium_2003,wu_valence-electron_2004,schowalter_ab_2005,kruse_determination_2006,deng_charge_2007}. Whenever an explicit comparison to the independent atom model was made, \emph{ab initio} values were consistently found to be lower and in better agreement with experiment. This tendency of the IAM to overestimate potential values was highlighted by very recent work on the mean inner potential of water~\cite{yesibolati_mean_2020}, which compared it to several experimental measurements (see Fig.~\ref{fig:water}).

\begin{figure}
    \centering
    \includegraphics[width=1\textwidth]{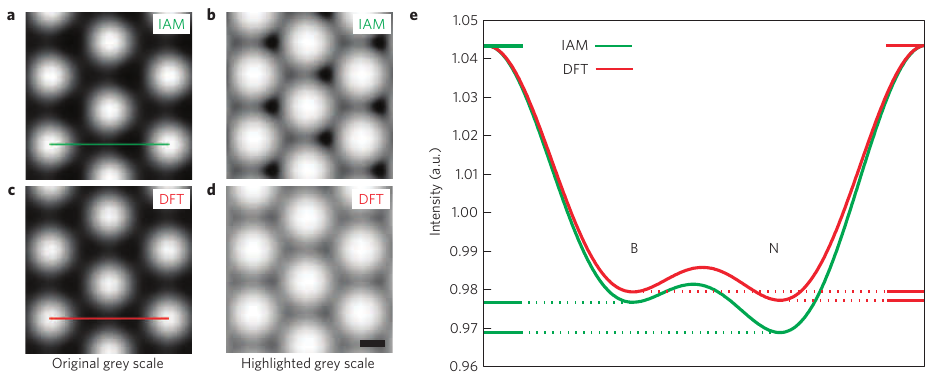}
    \caption{Comparison of high-resolution transmission electron microscopy (HRTEM) images simulated with the independent atom model (IAM; (a)-(b)) to those simulated with an electrostatic potential derived from density functional theory (DFT; (c)-(d)). The overlaid red and green lines in panels (a) and (c) correspond to the line profiles plotted in panel (e). Reproduced with permission from Ref.~\citenum{meyer_experimental_2011}, Copyright \copyright\ 2011, Springer Nature.}
    \label{fig:hBN}
\end{figure}

\begin{figure}
    \centering
    \includegraphics[width=0.65\textwidth]{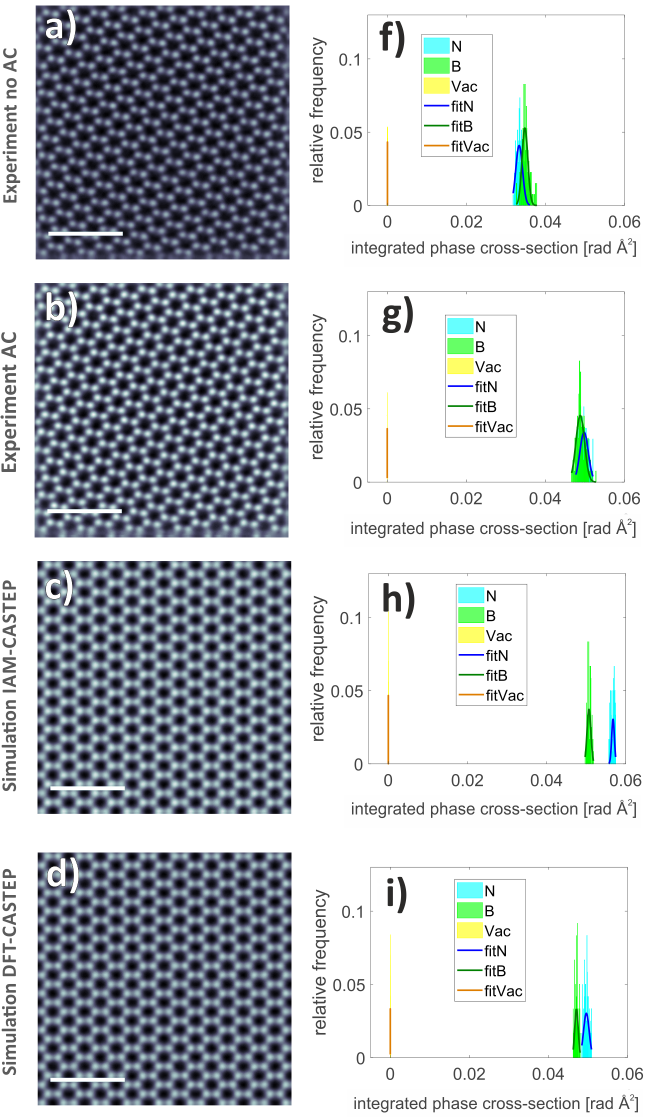}
    \caption{Ptychographic phase images of hBN from 4D-STEM experimental data (a) without and (b) with post-collection aberration correction (AC) compared to simulated data using (c) the IAM and (d) DFT. (f) to (i) display the corresponding histograms of the integrated phase cross-sections of N and B with respect to vacuum (Vac) with Gaussian fits, showing how DFT reduces the difference between the two sites in this ionic compound, bringing the results into decent agreement with experiment, although some discrepancy does remain. The scale bar in (a) to (d) is 1~nm. Adapted with permission from Ref.~\citenum{martinez2019direct}.}
    \label{fig:ptycho}
\end{figure}

In terms of spatially resolved studies, hexagonal boron nitride (hBN) has arguably been the clearest example of a system where valence bonding plays an important role for electron scattering. In contrast to independent B and N atoms, where the stronger nuclear scattering from the N sites would be expected to lead to their greater image intensity, the greater electronegativity of N in the bonded ionic compound leads to significant charge transfer from the B atoms, enhancing the screening of the N nucleus. This almost completely negates the contrast difference between the two sites in either HRTEM~\cite{meyer_experimental_2011} (see Fig.~\ref{fig:hBN}) and 4D-STEM ptychography~\cite{martinez2019direct} (based on the Wigner distribution deconvolution, see Fig.~\ref{fig:ptycho}), and can only be reproduced using an \emph{ab initio} electrostatic potential. This phenomenon was also responsible for the visibility of substitutional nitrogen sites in defocused HRTEM images of chemically doped graphene, again requiring an \emph{ab initio}-based potential to reproduce the experimental image contrast~\cite{meyer_experimental_2011,kurasch_simulation_2011}.

\begin{figure}[t]
    \centering
    \includegraphics[width=0.85\textwidth]{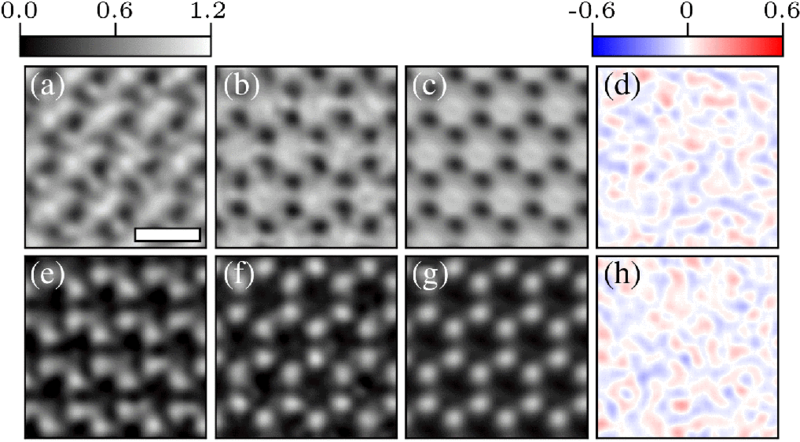}
    \caption{Comparison between experimental (off-axis holographic reconstruction with a 40\,mrad aperture applied to the sideband,  80\,keV) and \emph{ab initio} simulated~\cite{borghardt_quantitative_2017} wave functions for five-layer WSe$_2$. (a)–(d) Real parts and (e)–(h) imaginary parts of the electron wave functions. Experimental wave functions (a),(e) before and (b),(f) after residual aberration correction. (c),(g) Simulated wave function. (d),(h) Residual differences between simulation and experiment after aberration correction. The scale bar is 0.4\,nm. Reproduced with permission from Ref.~\citenum{winkler_absolute_2018}, Copyright \copyright\ (2018), American Physical Society.}
    \label{fig:WSe2}
\end{figure}

Another recent example is the comparison of atomically resolved off-axis holography of mono- and bilayer specimens of the transition metal dichalcogenide WSe$_2$~\cite{borghardt_quantitative_2017} to simulations. An \emph{ab initio} electrostatic potential was again required to reproduce electron-optical phase images of the material, though certain deviations remained and it is not clear whether these are due to limitations of the measurement or the model. In a subsequent study~\cite{winkler_absolute_2018}, the authors obtained an excellent match (the root-mean-square variation of the residual differences is close to the vacuum noise level) after approximating thermal diffuse scattering by smearing the \emph{ab initio} potential using a global Debye-Waller factor with $B = 0.003$\,nm$^2$ (Fig.~\ref{fig:WSe2}). Electron diffraction intensities are likewise highly sensitive to the inclusion of valence bonding, as was recently demonstrated for graphene and hBN, where only a scattering simulation with an \emph{ab initio} potential could correctly reproduce the intensity ratio of the first two diffraction orders~\cite{susi_efficient_2019}. While many of these examples are from two-dimensional materials, that is not due to any fundamental limitation; modern PAW-based DFT simulations can handle thousands of atoms, even when including an explicit description of van der Waals interactions~\cite{larsen_libvdwxc:_2017}.

\begin{figure}[t!]
    \centering
    \includegraphics[width=0.72\textwidth]{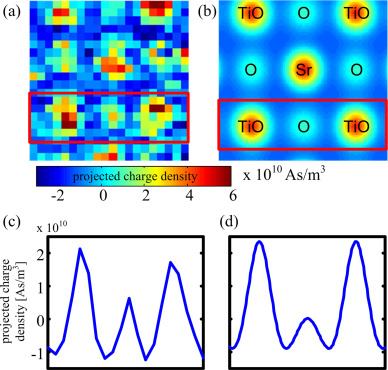}
    \caption{Projected charge density in the [100] projection of SrTiO$_3$ determined from average momentum transfers. (a) The probe-convolved charge density determined from a 20$\times$20 pixel momentum transfer map following scan noise correction. (b) Projected charge density from an \emph{ab initio} calculation. (c) and (d) show line scans averaged laterally across the red boxes in (a) and (b). Reproduced with permission from Ref.~\citenum{muller-caspary_measurement_2017}, Copyright \copyright\ (2017), Elsevier.}
    \label{fig:DPC}
\end{figure}

Finally, Fig.~\ref{fig:DPC} shows a recent example of STEM-based diffraction phase contrast (DPC) imaging in a bulk crystal. In their carefully systematic study, M{\"u}ller-Caspary mapped the projected charge density of SrTiO$_3$ in real space by recording a momentum-transfer map of 20$\times$20 pixels over several unit cells of the material using a pixelated detector. The data was corrected for scan noise and then converted into charge density via Gauss' flux theorem, and the resulting map compared to an \emph{ab initio} calculation. Good agreement was obtained between the measured and simulated data, even for the light oxygen columns that were not visible in either the bright-field or annular bright-field images from the same dataset.

\section{Outlook}
The rapid ongoing development of transmission electron microscopy instrumentation, imaging modalities, and analysis methods continues to expand the already impressive capabilities of the technique at a remarkable pace. At the same time, an intimate interplay with modeling is becoming ever more important, especially for techniques such as 4D-STEM and ptychography that aim at extracting novel physical information from each detected electron with the help of model-based reconstructions. Although these increase the computational cost and complexity of cutting-edge TEM work, such effort is still modest compared to the cost of density functional theory.

Advances in first-principles modeling codes and high-performance computing facilities are making ever-larger system sizes amenable for more physically accurate description. Further, emerging developments in machine learning and instrument automation can be expected to drive a trend towards closer software integration, for which new TEM simulation packages based on Python are particularly well suited.

While established simulation methods based on the independent atom model will continue to play an important role in routine work, we anticipate that the importance of accessing the full \emph{ab initio} electrostatic potential will grow ever more significant over time. Overall, the sustained increase in the power and versatility of transmission electron microscopy as a tool of choice for materials science and structural biology shows no sign of abating, and modern simulation methods and software tools are poised to make an important contribution to advancing both our scientific research and technological development.

\section*{Acknowledgments}
We thank our three expert referees for their remarkably detailed and useful feedback, as well as John Spence and Peter Rez for further helpful input. We acknowledge funding from the European Research Council (ERC) under the European Union’s Horizon 2020 research and innovation programme (J.M. and T.S. via Grant agreement No. 756277-ATMEN, and T.J.P. via 802123-HDEM).

%\bibliography{zotero.bib, mendeley.bib, mybibfile.bib}

\end{document}